\documentclass[aps,prl,floats, twocolumn,noshowpacs,superscriptaddress]{revtex4}

\usepackage{graphicx,epsfig}
\usepackage{times}
\usepackage{graphics,dcolumn,bm,epic, eepic,fleqn,float}
\usepackage{amssymb,amsmath,multirow,rotate,color}

\def\Erdos{Erd\"os}
\begin{document}

\title{Cooperation in changing environments:\\ Irreversibility in the transition to cooperation in complex networks}


\author{C. Gracia-L\'azaro}

\affiliation{Institute for Biocomputation and Physics of Complex
Systems (BIFI), University of Zaragoza, Zaragoza 50009, Spain}

\author{L. M. Flor\'{\i}a}


\affiliation{Institute for Biocomputation and Physics of Complex
Systems (BIFI), University of Zaragoza, Zaragoza 50009, Spain}

\affiliation{Departamento de F\'{\i}sica de la Materia Condensada,
University of Zaragoza, Zaragoza E-50009, Spain}

\author{J. G\'{o}mez-Garde\~{n}es}

\affiliation{Institute for Biocomputation and Physics of Complex
Systems (BIFI), University of Zaragoza, Zaragoza 50009, Spain}

\affiliation{Departamento de F\'{\i}sica de la Materia Condensada,
University of Zaragoza, Zaragoza E-50009, Spain}

\author{Y. Moreno}


\affiliation{Institute for Biocomputation and Physics of Complex
Systems (BIFI), University of Zaragoza, Zaragoza 50009, Spain}

\affiliation{Departamento de F\'{\i}sica Te\'orica. University of
Zaragoza, Zaragoza E-50009, Spain}

\begin{abstract}
In the framework of the evolutionary dynamics of the Prisoner's Dilemma game on complex networks, we investigate the possibility 
that the average level of cooperation shows hysteresis under quasi-static variations of a model parameter (the ``temptation to defect''). 
Under the ``discrete replicator'' strategy updating rule, for both \Erdos -R\'enyi and Barab\'asi-Albert graphs we observe cooperation hysteresis 
cycles provided one reaches tipping point values of the parameter; otherwise, perfect reversibility is obtained. The selective fixation of cooperation
at certain nodes and its organization in cooperator clusters, that are surrounded by fluctuating strategists, allows the rationalization of the 
``lagging behind'' behavior observed.
\end{abstract}



\maketitle

\section{Introduction}
\label{sct:Introduction}

Evolutionary game theory \cite{nowakbook} provides with an elegant mathematical description of how 
Darwinian natural selection among strategies (representing phenotypes) takes place when the reproductive 
success of individuals (and thus the future abundance of phenotypes) depends on the current
strategic composition of the population (frequency-dependent fitness) \cite{hofbauer,gintis,hofbauerbull}. 
One of the most studied challenges to the explanatory power of evolutionary game dynamics is the
understanding of the  evolutionary survival of cooperative behavior among unrelated individuals that is 
observed even when selfish actions provide higher short-term benefits. 

Perhaps the best suited mathematical model to describe the puzzle of how cooperation survives is the Prisoner's Dilemma (PD) game. This game, originally introduced by M.M. Flood and G.J. Savage \cite{rand}, is a two-players-two-strategies game, in which each player chooses one of the two available strategies: cooperation (C) or defection (D). A cooperator earns $R$ (reward) when playing with a cooperator, and a payoff $S$ (sucker's payoff) when playing with a defector, while a defector earns $P$ (punishment) when playing with a defector, and $T$ (temptation to defect) against a cooperator. When the ordering of the payoffs is $T>R>P>S$, the game is strictly speaking a PD. When $T>R>S>P$ the situation is called Snowdrift Game (SG), also known as Hawks-and-Doves or Chicken \cite{maynard}. In this work we focus on a variant of the PD game called weak Prisoner's Dilemma, placed in its boundary with respect to SG, that is $T>R>P=S$. In a PD (including weak variant), whatever the opponent's strategy is, the payoff is never higher for cooperation, and a rational agent should choose defection. Still, two cooperator agents receive higher payoff ($2R$) than two defector ones ($2P$), which leads to social dilemma. 

On the other hand, recent discoveries on the interaction architecture of biological, technological and social systems have shown that this structure has important consequences for their dynamical behavior and their associated critical phenomena \cite{rev:albert,rev:newman}. In particular, the dynamical features observed for heterogeneous, scale-free (SF) networks are often different from those for homogeneous networks \cite{rev:bocc,rev:doro}. This difference is rooted in the presence of highly connected nodes. 
Motivated by the aforementioned results, studies of evolutionary game theory models
on heterogeneous networks have attracted a great attention in the last few years \cite{rev:szabo,rev:roca,rev:jrsi}. In particular, issues concerning the role that social structure plays in the success of cooperative behavior have been extensively studied in the context of
the PD game \cite{pacheco1,pacheco2,pacheco3,pacheco4,campillo,poncela,jtb,SzolnokiPhA2008,floria}. The results obtained indicate that the heterogeneous connectivity patterns characteristic of SF networks might support cooperation, although it has been recently reported that when humans play a PD game, networks do not play a role \cite{pnas}. The promotion of cooperation when the payoff differences determine the evolution of strategies is understood by the presence of highly connected nodes (the hubs) whose large number of connections provides them with high benefits and then a high imitation power. Thus, when playing as cooperators neighboring agents easily imitate their cooperative strategy. This imitation reinforces the benefits accumulated by the cooperator hubs thus creating a positive feedback mechanism that enhances the cooperation across the whole system.

The asymptotic levels of average cooperation on random networks (as function of game's parameters) were determined to be reasonably robust versus variation of random initial conditions with different initial proportion of cooperators \cite{poncela}. Though this might seem to suggest that, under slow (compared to evolutionary scales) parameter variation, the cooperation levels reached on a network are independent of the particular history of parameters' change, it is unclear that it should be so, because of the likelihood of multiplicity of microscopic asymptotic states, given the disordered nature of the social contacts structure, that could produce hysteresis-like behaviors. We address here the question about the possibility of hysteresis of the cooperation on complex network under quasi-static variation of game's parameters.

\section{The model}

Provided the relative selective advantage among two
individuals depends on their payoff's difference (see below), we can normalize without
loss of generality the pay-off matrix taking $R=1$ and fix the punishment $P=0$.
We will also consider here that the sucker's payoff is $S=0$, namely, the weak prisoner's dilemma, so that both strategies perform equally against a defector. 
Then only one parameter, the ``temptation to defect'', $T=b>1$ is a system variable.

In this study we implement the following replication mechanism: At each time step, each agent
$i$ plays once with each one of its neighbors ({\em i.e.} agents connected to $i$)
and accumulates the obtained payoffs, $P_i$. After that, the individuals, $i$, update
synchronously their strategies choosing a neighbor $j$ at random, and comparing their
respective payoffs $P_i$ and $P_j$. If $P_i\geq P_j$, nothing happens and $i$ preserves
its strategy. Otherwise, if $P_j>P_i$, $i$ adopts the strategy of its neighbor $j$
with probability $\Pi_{ji}=\eta(P_j-P_i)$. Next, all payoffs are reset to zero. Here,
$\eta$ is a positive real number, related to the characteristic inverse time scale: the
larger it is, the faster evolution takes place.  We consider that players and
connections between them are given by a fixed graph where agents are represented by
nodes, and a link between nodes indicates that they interact. We choose here the maximum
value of $\eta$ that preserves the probabilistic character of $\Pi_{ji}$, that is,
$\eta = (\mbox{max}\{k_i,k_j\} b)^{-1}$, where $k_i$ is the number of neighbors of agent
$i$ (connectivity or degree). This choice, introduced in \cite{pacheco2}, slows down the
invasion processes from or to highly connected nodes (hubs), with respect to the rate of
invasion processes between poorly connected nodes.

Our aim is the study of the reversible (or irreversible) character of cooperation
level $c$ under the variation of the temptation to defect parameter $b$,
where $c$ is defined as the number of cooperator nodes divided by the total
population $c=N_c/N$. In order to study the system's behavior, we choose an
initial value of $b=b_0$ such that the asymptotic cooperation value $c$ is
close to a half: $c(b_0)\simeq0.5$. Once the system has reached a stationary
state, we decrease $b$ in a quasi-static way, that is, in steps $\Delta b<0$ small
enough to ensure that the system remains very close to equilibrium. Along this
process, we compute the stationary value of cooperation $c(b)$ for each value
of $b$. To avoid getting stuck in the absorbing states we deal with large
enough network sizes ($N>10^5$), considering that fluctuations decrease
according to the square root of the system size. Once the system has almost
reached the absorbing state $c=0$, we reverse the sign of the increase in
$b$, {\em i.e.} $\Delta b>0$, to almost reach the other absorbing state $c = 1$, and then again
decrease $b$ to complete the cycle. To study the influence of network topology
in the reversibility of the process, we consider three different network
models: Random Regular Graphs (RRG), \Erdos- R\'enyi and Scale-free networks, 
though we will only show here the results for the last two types of graphs. RRG 
(\emph{i.e.}, random networks with fixed degree $k$, which means that every node
has the same number of neighbors) always show complete reversibility. This seems to imply that, although randomness is present in RRG, 
a non-zero variance in the degree distribution is a necessary condition for the 
observation of irreversible behavior. 


\section{\Erdos-R\'enyi networks}
\Erdos-R\'enyi (ER) networks are random graphs characterized by a binomial degree
distribution (or a Poisson distribution for large networks). To study reversibility, we have performed numerical
simulations in $10^3$ independent networks of size $N=1.2\times 10^ 5$ generated
through \Erdos-R\'enyi algorithm. For reduced cycles, that is,
when the return points are far from absorbing states
($1-N_c(b_{min})\gg 1$, $N_c(b_{max})\gg 1$) the processes
are reversible and the level of cooperation is
independent of the sign of the increase in $b$. Nevertheless, when return
points are close enough to the absorbing states
($c(b_{min})\approx 1, c(b_{max})\approx 0$), ER networks exhibit irreversibility to a large extent. In fact, once the level of cooperation reaches a tipping point,
all processes are irreversible. In particular, there is a strong resilience
of cooperation (defection) when increasing (decreasing) the value of $b$.
However, the backward and forward transition curves are identical for
intermediate values of cooperation. The proximity $\epsilon$ of the tipping
points $c(b_{min}),c(b_{max})$ to the absorbent states in both ends of the cycle
turns out to be similar: $1-c(b_{min})=\epsilon \approx c(b_{max})$ and, for
the networks size used, it takes on the value $\epsilon \approx 2 \times 10^{-3}$.

\begin{figure}
\begin{center}
\epsfig{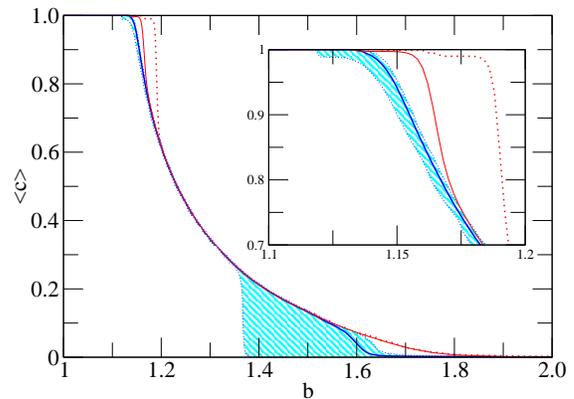}
\end{center}
\caption{(color online) Cooperation level $\langle c \rangle$ versus the
temptation to defect $b$  averaged over $10^ 3$ ER networks (solid lines)
and envelopes (pointed lines). Red lines represent semicycles with
increasing $b$ and blue lines represent semicycles with decreasing $b$.
The network size is $N=1.2\times 10^5$. See the text for further details.
}
\label{cyclesER}
\end{figure} 

As a result, once the population has reached a cooperation level above (below)
a tipping point, the system shows a reticence to retrieve the past level of
cooperation when the parameter $b$ increases (decreases). This phenomenon is
independent of the particular ER network, being observed in all network realizations.
Figure \ref{cyclesER} shows the level of cooperation $\langle c\rangle$ versus
the temptation to defect $b$, averaged over $10^ 3$ realizations in distinct
ER networks. Different realizations show different $b$-increasing and $b$-decreasing 
curves, whose envelopes are depicted as dotted lines in Figure \ref{cyclesER}. 
Remarkably, the dispersion of the different curves is much larger for the $b$-decreasing 
direction.

\section{Scale-free networks}
 
Scale-free (SF) networks are graphs whose degree distribution $P(k)$ follows
a power law, that is, $P(k) \sim ck^{-\gamma} $. We ran simulations in $5\times 10^3$
independent networks of size $N=1.2\times 10^ 5$ generated through the Barab\'asi-Albert 
algorithm. Although most of the SF networks show nearly reversible behavior, around $5\%$ 
of networks show a strong hysteresis. Nevertheless, irreversibility in SF networks should not
be considered as a rare event: Increasing the network size increases the proportion of networks 
that show irreversible behavior. The explanation for this fact is that the use of larger networks 
allows to approach closer the absorbing states $c=0,1$ without getting stuck in them.
Based on this argument, we have separated realizations showing a reversible behavior from
irreversible ones. In these latter cases, hysteresis shows up only for low values of $b$; in other 
words, when cooperation is very small, backward and forward $c(b)$ curves are almost
identical. Moreover, the behavior of the system in $b$-increasing semi-cycles is always
similar, the cooperation level $c(b)$ taking approximately the same value in
all realizations, regardless of whether they are reversible or irreversible. On the contrary, $c(b)$ curves
are different for different (irreversible) realizations in $b$-decreasing semi-cycles, and show 
a substantially larger dispersion than those of ER networks. 

\begin{figure}
\begin{center}
\epsfig{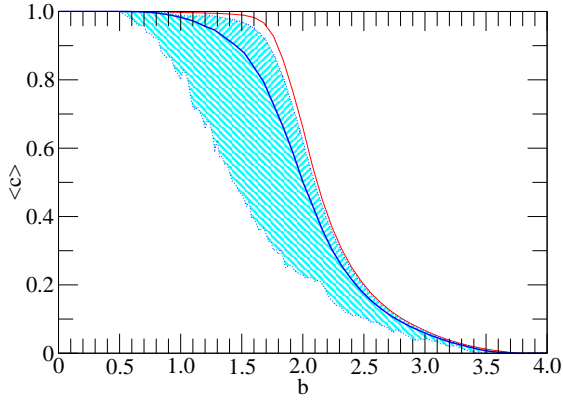}
\end{center}
\caption{(color online) Cooperation level $\langle c\rangle$ versus the temptation
to defect $b$ averaged over $100$ SF networks (solid lines) and envelopes (dotted
lines). Red lines represent semi-cycles with increasing $b$ and blue lines represent
semi-cycles with decreasing $b$. Only irreversible realizations are shown.
The network size is $N=1.2\times 10^5$. See the text for further details.
}
\label{cyclesSF}
\end{figure}

The results of the average cooperation level $\langle c \rangle$ as a function of
the temptation to defect $b$, for SF networks showing irreversible behavior, are 
presented in figure \ref{cyclesSF}. The return points
$b_{min}, b_{max}$ were chosen such that $c(b_{max})=1-c(b_{min})=\epsilon$, for
a value of $\epsilon=10^{-3}$. Note that, despite the small value of $\epsilon$, the
network size $N$ is large enough so as to ensure that we are not dealing with pathological
cases, since a value $c=0.001$ involves a number of cooperators $N_c=120$. In
the same way, $c=0.999$ implies 120 defector nodes. As shown in envelopes
(dotted lines), the degree of irreversibility varies greatly from one realization to another.
Specifically, irreversibility depends on the particular network, since for a given 
network repeated cycles share approximately the same $ c (b)$ curves for a given  
(forward or backward) direction. A most remarkable feature of the irreversibility in 
SF networks is that, for irreversible network realizations, the value of the temptation to 
defect needed to reach a cooperation level of $c=10^{-3}$ is $b_{min}<1$, 
that is to say, outside the PD game range.

\begin{figure}
\begin{center}
\epsfig{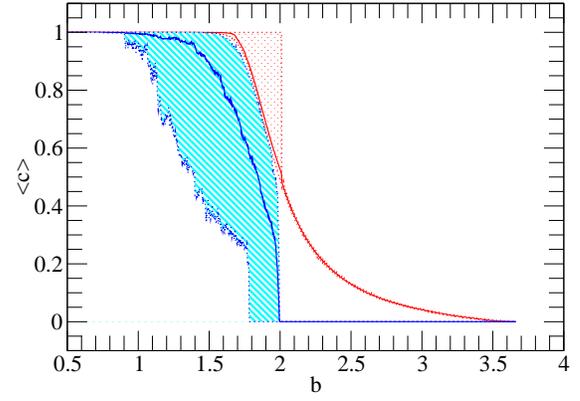}
\end{center}
\caption{(color online) Cooperation level $\langle c\rangle$ versus the temptation
to defect $b$ averaged over $100$ SF networks (solid lines) and envelopes (dotted
lines). Red lines represent semicycles with increasing $b$ and blue lines represent
semicycles with decreasing $b$. In this case, a constraint has been considered: $N_c>\mu$ and
$N-N_c>\mu$. All the realizations have been taken into account.
The network size is $N=1.2\times 10^5$ and $\mu=10$. See the text for further details.
}
\label{SFconstraint}
\end{figure}

In order to validate (or refute) the hypothesis that
most of the studied SF networks do not show irreversibility because the return points are not
close enough to the tipping points, we have addressed the problem through a modification
of the model: to avoid getting stuck in absorbing states ($c=0,1$), now we add a constraint which
keeps the minimum number of cooperators and defectors above a given threshold $\mu$. In particular,
we add the condition that a node can switch its strategy to
defection (\textit{resp.}, cooperation) only if $N_c>\mu$ (\textit{resp.}, $N-N_c>\mu$). Figure \ref{SFconstraint}
shows the cooperation level $c$ as a function of
the temptation to defect $b$, averaged over 100 different SF networks. The values of return points $b_{min}, b_{max}$ have been
chosen such that $N_c(b_{max})=N-N_c(b_{min})=\mu$, for a value of $\mu=20$. In this case, all studied
networks show a strong hysteresis indicating that, once the return points have been chosen close enough to the absorbing states,
the process is always irreversible.

\section{Microscopic roots}

\begin{figure}
\begin{center}
\epsfig{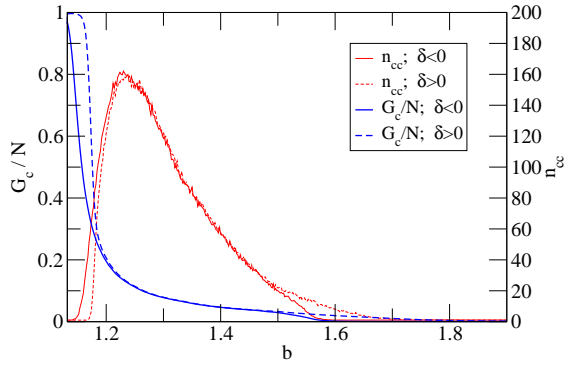}
\end{center}
\caption{(color online) 
Number $n_{cc}$ of cooperator clusters (blue, thick lines) and relative size of main
cooperator cluster $G_c/N$ (red, thin lines) in ER networks.
Solid lines represent $b$-decreasing half-cycles and dashed lines
represent $b$-increasing half-cycles. The system size is $N=1.2\times 10^ 5$.
We have averaged over 50 simulations.
}
\label{clustersER}
\end{figure}

Previous studies \cite{campillo,poncela,jtb,floria} have shown that, in the 
asymptotic states of the evolutionary dynamics of the PD game, under the updating rule explained above (discrete replicator rule), the network 
is generically partitioned into three sets of nodes: Pure cooperators (nodes where 
cooperation has reached fixation), pure defectors, and fluctuating strategists 
(nodes where fixation is impossible so that defection and cooperation alternate 
forever). Pure cooperators resist invasion by grouping together in cooperator clusters, 
each of these connected subgraphs keeping around it a cloud of fluctuating strategists.
The basis for an understanding of the irreversible behavior in ER networks is 
found by looking along both ($b$-increasing and $b$-decreasing) branches 
at the details of this microscopic organization of cooperation. In particular, in 
what follows we pay attention to the number and size of pure cooperator clusters 
as a function of $b$. Figure \ref{clustersER} shows the averaged relative size
$\langle G_c/N \rangle $ of the largest cooperator cluster, and the average 
$\langle n_{cc} \rangle$ of the number of cooperator clusters versus 
the temptation to defect $b$, in both semi-cycles for ER networks.

Let us first analyze the $b$-increasing semi-cycle. In typical configurations near the 
absorbing state $c = 1$,  pure cooperators percolate the network forming 
a giant cooperator cluster whose averaged relative size $\langle G_c/N \rangle \simeq 1$. 
As the temptation to defect $b$ increases, starting from such configurations, the 
existence of a single very large cluster of pure cooperators allows initially for 
a very efficient resilience to invasion by defectors until a value of $b\simeq 1.16$ 
is reached. From there on, invasion processes are greatly enhanced, thus 
inducing the fragmentation of the largest cluster: $\langle G_c/N \rangle$ decreases 
quickly, the largest cluster give rise to an increasing number $n_{cc}$ of small 
clusters of pure cooperators. At $b\simeq 1.23$, this quantity reaches its maximum value 
$n_{cc}\simeq 160$ and the largest cluster size has been reduced to 
$\langle G_c/N \rangle \simeq 0.15$. Further increase of $b$ reduces both the 
number of pure cooperator clusters and the size of the largest one: At $b\simeq 1.8$ 
basically only the largest cluster remains with a very small size which keeps 
decreasing further beyond the tipping point (typically found at $b\geq 2$).

Now we analyze the $b$-decreasing semicycle. Back from the typical configuration 
reached past the tipping point near the absorbing state $c = 0$, when 
decreasing the temptation value $b$ the very small size of the remaining pure cooperator 
cluster cannot benefit ({\em i.e.}, enlarge its size) enough from the cooperative fluctuations 
nearby; correspondingly the level of cooperation $\langle c\rangle$ remains well below 
the values observed for the $b$-increasing branch. It is not until a value of $b\simeq 1.6$ 
is reached, that $\langle G_c/N \rangle$ starts a significant increase. Simultaneously, 
some cooperative fluctuations in the cloud of fluctuating agents form independent (separated) small 
cooperator clusters, so that $n_{cc}$ also starts to significantly deviate from zero. At 
around $b\simeq 1.5$ both $\langle G_c/N \rangle$ and $n_{cc}$ (as well as the average 
level of cooperation $\langle c\rangle$) show already values that are very close to those 
exhibited by the $b$-increasing branch. However, once reached the value $b\simeq 1.23$, 
where $n_{cc}$ has its maximum value (and, as explained previously, the 
fragmentation of the largest cluster of pure cooperators reached an end in the $b$-increasing 
branch), a further decrease in $b$ leads to an increase of $\langle G_c/N \rangle$, 
and a concomitant decrease of $n_{cc}$ due to the connection of small cooperator clusters 
to the largest one. These processes take place at a slower pace than the corresponding 
fragmentation occurring for the $b$-increasing branch. The consequence is that the 
values of the cooperation level in this range of $b$ for the $b$-decreasing branch, are 
significantly lower than those for the $b$-increasing semi-cycle. Note that though the values 
of $\langle G_c/N \rangle$, $n_{cc}$, and $\langle c\rangle$ in the range of intermediate 
$1.23 \leq b \leq 1.5$ values are very similar in both branches, the system keeps memory 
of the path followed, demonstrating the importance of the particular topological details of 
the organization of cooperator clusters.

\begin{figure}
\begin{center}
\epsfig{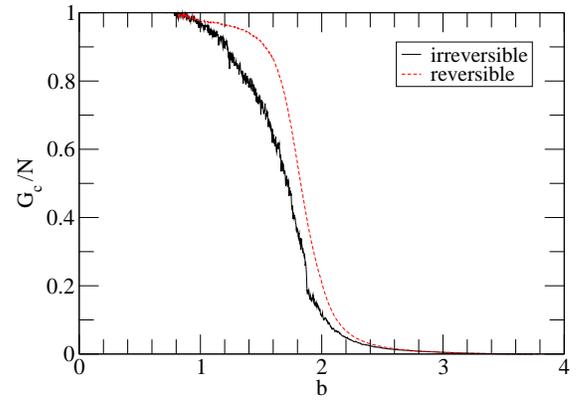}
\end{center}
\caption{(color online) Relative size of the main cooperator cluster $G_c/N$ for reversible
processes (dashed line, red) and irreversible ones (solid line, black) in the
$b$-decreasing semicycle ($\Delta b < 0$) on SF networks. Averaged over the $100$ different networks 
studied that show irreversible behavior. The system size is $N=1.2\times 10^ 5$.
}
\label{nccSF}
\end{figure}

A significant difference, regarding the microscopic organization of cooperation, between 
ER and SF networks, is the observation first reported in \cite{campillo} that for SF networks
pure cooperators group together in a single cluster, while in ER networks they are 
disaggregated into several cooperator clusters for generic values of $b$. In our simulations 
here we are using network sizes that are larger than those used in \cite{campillo} by a factor 
of $30$, and for SF networks we have observed nodes that, though being isolated from the 
main cooperator cluster, remain cooperators during observational time scales. Strictly speaking 
they are not pure cooperators, for the probability of invasion by the defective strategy is not 
strictly zero (in all the cases analyzed), though it turns out to be exceedingly small, due to the 
large connectivity (degree) of these nodes. These quasi-pure cooperators appear in both branches; 
also, they are present in reversible paths, i.e., those where return points are chosen before reaching 
tipping points. For a network size of $N=1.2\times 10^ 5$ its number is never larger than $8$ for 
$b$-increasing branches, and not larger than $14$ for $b$-decreasing branches. Their contribution 
both direct and indirect (through the cloud of fluctuating strategists each one keeps nearby) to the 
level $c$ of cooperation can be considered as negligible. Still one cannot discard {\em a priori} an 
eventual role they might play in the reshaping of the main cooperator cluster during the hysteresis 
cycle of particular irreversible realizations. 

In figure \ref{nccSF} we plot the relative size of the cooperator cluster $\langle G_c/N \rangle$
averaged over $100$ irreversible realizations for both forward and backward branches of the 
cycle. Contrary to what happens for ER networks at high values of the temptation to defect,
when starting to decrease it from $b_{min}$, the size of the cooperator cluster
in SF networks initially follows very closely the values of the forward branch until $b \simeq 2.5$. 
However, significant differences in the average cooperation value $\langle c\rangle$ (see figure 
\ref{cyclesSF}) are already noticeable from $b\simeq 3$, indicating that the contribution from the 
cloud of fluctuating strategies is lower for the backward branch. When further decreasing $b$ 
down from $b \simeq 2.5$, the averaged size of the cooperator cluster takes on values 
progressively lower than in the $b$-increasing branch. This agrees nicely with the observation 
just made in previously regarding the cloud of fluctuating strategies, for the growth of the 
cooperator cluster originates from the cooperative fluctuations in its frontier, and thus the strength 
of these fluctuations determines the pace of the cluster size growth. The difference between 
forward and backward branches persists down to the tipping point, which somewhat surprisingly 
occurs for values of $b$ outside the PD game range.

\section{Conclusions}

In this paper, we have studied the evolutionary dynamics of a classical PD game on top of complex homogeneous and heterogeneous networks. Although there are many results available in the literature, here we have addressed a problem that can be of interest in real settings. Admittedly, to the best of our knowledge, all known results have dealt with situations in which the average levels of cooperation reported are obtained by averaging over different initial conditions for fixed values of the model parameters. However, real system are continuously evolving from a given micro-state $-$ characterized by a distribution of strategists $-$ to another one driven by both the internal dynamics (the evolutionary rules) and the external conditions. The latter can be ultimately modeled as if the entries of the payoff matrix were changed. For instance, in some conditions, the pressure towards cooperation could be greater than in other situations or the other way around. This means that parameters such as the temptation to defect could change continuously and that such a change takes place without reseting the system, i.e., the initial condition is the previous steady state of the system.

Our results show that the generic existence of different stationary strategic configurations for the evolutionary dynamics of the PD game 
on networks (under the discrete replicator strategy updating rule) allows for a hysteresis-like behavior of the average cooperation 
level when the model parameter is quasi-statically varied. However, our results on random complex networks indicate
that the observation of hysteresis is greatly conditioned by the range of variation of the parameter. Unless the  
temptation to defect $b$ is brought to a large (resp. small) enough value, so as the cooperation level reaches an almost zero 
(resp. unity) value, the behavior is perfectly reversible.

\section{Acknowledgement}
\label{sct:Acknowledgement}
We thank Jos\'e Cuesta for many helpful comments and discussions. This work has been partially supported by MINECO through Grant FIS2011-25167; Comunidad de Arag\'on (Spain) through a grant to the group FENOL and by the EC FET-Proactive Project MULTIPLEX (grant 317532).


\begin{thebibliography}{99}
\expandafter\ifx\csname
natexlab\endcsname\relax\def\natexlab#1{#1}\fi
\providecommand{\bibinfo}[2]{#2} \ifx\xfnm\relax
\def\xfnm[#1]{\unskip,\space#1}\fi


\bibitem{nowakbook} M.A. Nowak, Evolutionary Dynamics: Exploring the Equations of Life, Harvard University Press, Cambridge, 2006.

\bibitem{hofbauer} J. Hofbauer, K. Sigmund, Evolutionary Games and Population Dynamics, Cambrige University Press, Cambridge, 1998.

\bibitem{gintis} H. Gintis, Game Theory Evolving, Princeton University Press, Princeton, 2000.

\bibitem{hofbauerbull} J. Hofbauer, K. Sigmund, Evolutionary game dynamics, Bull. Am. Math. Soc. 40 (2003) 479–519.

\bibitem{rand} M. M. Flood, L.J. Savage, A Game Theoretic Study of the Tactics of Area Defense, RAND Research Memorandum 51 (1948).

\bibitem{maynard} J. Maynard Smith, G.R. Price, The Logic of animal Conflict, Nature 246 (1973) 15-18. 

\bibitem{rev:albert} R. Albert, and A.-L. Barab\'asi, Statistical mechanics of complex networks, Rev. Mod. Phys. 74 (2002) 47-97.

\bibitem{rev:newman} M.E.J. Newman, The structure and function of complex networks SIAM Rev. 45 (2003) 167-256.

\bibitem{rev:bocc} S. Boccaletti, V. Latora, Y. Moreno, M. Chavez and, D.-U. Hwang, 
Complex networks: Structure and dynamics, Phys. Rep. 424 (2006) 175-308.

\bibitem{rev:doro} S.N. Dorogovtsev, A.V. Goltsev, and J.F.F. Mendes, Critical phenomena in complex network,
Rev Mod. Phys. 80 (2008) 1275-1335.

\bibitem{rev:szabo} G. Szab\'o, G. Fath, Evolutionary games on graphs, Phys. Rep. 446 (2007) 97-216.

\bibitem{rev:roca} C.P. Roca, J.A. Cuesta, A. S\'anchez, Evolutionary game theory: Temporal and spatial
effects beyond replicator dynamics, Phys. Life Rev. 6 (2009) 208-249.


\bibitem{rev:jrsi} M. Perc, J. G\'omez-Garde\~nes, A. Szolnoki, L.M. Flor\'{\i}a, Y. Moreno,
Evolutionary dynamics of group interactions on structured populations – A review, J. Roy. Soc. Interface 10 (2013) 20120997.

\bibitem{pacheco1} F.C. Santos, J. M. Pacheco, Scale-free networks provide a unifying framework for the emergence
of cooperation, Phys. Rev. Lett 95 (2005) 098104.

\bibitem{pacheco2} F. C. Santos,  J. M. Pacheco, A new route to the evolution of cooperation, J. Evol. Biol. 19 (2006) 726-733. 

\bibitem{pacheco3} F.C. Santos, J. F. Rodrigues, J.M. Pacheco, Graph topology plays a determinant role in
the evolution of cooperation, Proc. Biol. Sci. 273 (2006) 51-55.

\bibitem{pacheco4} F. C. Santos J. M. Pacheco, T. Lenaerts, Evolutionary dynamics of social dilemmas in
structured heterogeneous populations, Proc. Natl. Acad. Sci. USA 103 (2006) 3490-3494.

\bibitem{campillo} J. G\'omez-Gardenes et al., Dynamical Organization of Cooperation in Complex Topologies, Phys. Rev. Lett. 98
(2007) 108103.

\bibitem{poncela} J. Poncela, J. G\'omez-Gardenes, L.M. Flor\'{\i}a, Y Moreno, Robustness of
cooperation in the evolutionary prisoner's dilemma on complex networks, New J. Phys. 9 (2007) 184-189.

\bibitem{jtb} J. G\'omez-Gardenes et al. Natural selection of cooperation and degree
hierarchy in heterogeneous populations, J. Theor. Biol. 253 (2008) 296-301.

\bibitem{SzolnokiPhA2008} A. Szolnoki, M. Perc, Z. Danku, Towards effective payoffs in the prisoner’s
dilemma game on scale-free networks, Physica A 387 (2008) 2075-2082.


\bibitem{floria} L. M. Flor\'{\i}a et al. Social network reciprocity as a phase transition in evolutionary cooperation,
Phys. Rev. E 79 (2009) 026106.

\bibitem{pnas} C. Gracia-L\'azaro et al. Heterogeneous networks do not promote cooperation when humans play a Prisoner's Dilemma, Proc. Natl. Acad. Sci. USA 109 (2012) 12922. 

\bibitem{RMPsoc} C. Castellano, S. Fortunato, and V. Loreto, Statistical physics of social dynamics,
Rev. Mod. Phys. 81, (2009) 591-646.


\end{thebibliography}

\end{document}